\documentclass[showpacs,]{revtex4}
\usepackage{bm}
\usepackage{amsmath}
\usepackage{epsfig}
\usepackage{graphicx}
\usepackage{color}

\begin{document}
\title{Scalar field in nuclear matter: the roles of spontaneous chiral symmetry breaking and nucleon structure}
\author{G. Chanfray} 
\affiliation{IPN Lyon, Universit\'e de Lyon, Univ.  Lyon 1, 
 CNRS/IN2P3, UMR5822, F-69622 Villeurbanne Cedex}
\author{M. Ericson} 
\affiliation{IPN Lyon, Universit\'e de Lyon, Univ.  Lyon 1, 
 CNRS/IN2P3, UMR5822, F-69622 Villeurbanne Cedex}
\affiliation{Theory division, CERN, CH-12111 Geneva }
\begin{abstract}
Chiral theories with spontaneous symmetry breaking such as the Nambu-Jona-Lasinio (NJL) model lead to the existence of a scalar mode. We present in a detailed manner  how the corresponding low momentum effective lagrangian involving the scalar field can be constructed starting from the NJL model.  We discuss the relevance of the scalar mode for the problem of the nuclear binding and saturation. We show that it depends on the nucleon mass origin with two extreme cases. If this origin is entirely due to confinement the coupling of this mode to the nucleons vanishes, making it irrelevant for the nuclear binding problem. If instead it is entirely due to spontaneous symmetry breaking it couples to the nucleons
 but nuclear matter collapses. It is only in the case of a mixed origin with spontaneous breaking that nuclear matter can be stable and reach saturation. We describe models of nucleon structure where this balance is achieved. We also show how chiral constraints and confinement modify the QCD sum rules for the mass evolution in nuclear matter.    
\end{abstract}

\pacs{24.85.+p 11.30.Rd 12.40.Yx 13.75.Cs 21.30.-x} 
\maketitle
\section{Introduction}

The relation between the fundamental properties of low energy QCD, namely chiral symmetry and confinement, and the nuclear many-body problem is one of the most challenging aspect  of present day nuclear physics.   One  question is how the interplay between 
chiral symmetry and confinement in the nucleon structure manifests itself  in  the nuclear many-body problem. In a set of recent papers \cite{CE05, CE07, MC08} we have   associated the mean-field nuclear attraction  with the in-medium modification  of a (chiral invariant) background scalar field which reflects part of the  evolution of the chiral quark condensate. In this framework the nuclear medium can be seen as a shifted QCD vacuum. Nuclear stability is ensured with the phenomenological incorporation  of the nucleon  response to this  scalar field. This  response  depends on the quark confinement mechanism inside the nucleon.  This  framework has been implemented in nuclear matter calculation at the Hartree level \cite{CE05}. In a subsequent work  \cite{CE07} we also incorporated non relativistically the pion loop correlation energy. A full relativistic Hartree-Fock (RHF) calculation was then done in \cite{MC08} allowing to reproduce also the asymmetry properties of nuclear matter.

The aim of this paper is to discuss the foundations of this picture and  the nature of this scalar background field. Although the concept of a  scalar  field  has been widely used for nuclear matter studies \cite{SW86}  its precise origin or meaning  is still a controversial subject. The problem is that there is no sharp scalar resonance which would lead to a simple scalar particle exchange. In our approach instead we stress the chiral aspect of the problem. As soon as we start from a model which gives a correct description of chiral symmetry breaking in the QCD vacuum such as the Nambu-Jona-Lasinio model (NJL), the emergence of a scalar field linked to the quark condensate cannot be avoided. This is by construction a low momentum concept which does not imply the existence of a sharp scalar meson if the effect of confinement is taken into account. Indeed it has been demonstrated by Celenza et al \cite{CSWSX95,CWS01} that the inclusion of a confining interaction on top of the NJL model pushes the $q\bar q$ scalar state, located originally at twice the constituent quark mass,  well above one GeV. As for the $f_0(600)$
 it appears as a broad  resonance $\pi\pi$ resonance which has no direct relation with the scalar field.  The explicit construction of the scalar field can be done using  a bozonization technique based on a derivative expansion valid et low (space-like)
 momenta.  The corresponding ``scalar mass'', which   is around twice the constituent quark mass,  is in a fact a low 
momentum parameter related to the inverse of the vacuum scalar susceptibility. 
We remind for completeness that, according to \cite{CSWSX95,CWS01}, the confining interaction has little influence on the low momentum parameters entering the effective lagrangian. 

A priori the range of this mass and the magnitude of the scalar coupling   to the nucleon make it relevant for nuclear physics. The real question for  this relevance  is  intimately related to the problem of the structure of the nucleon and the origin of its mass. The respective roles of spontaneous symmetry breaking and confinement in the generation of this mass are indeed crucial. Confinement has little effect
on the low momentum parameters but it  leads to the concept of  a nucleonic response to the scalar field, as was originally introduced by P. Guichon \cite{G88}. Without it, ({\it i.e.}, in the pure NJL model), nuclear matter would not be stable and would collapse \cite{BT01}, due to attractive three-body forces (tadpole diagram). On the other hand if the nucleon mass were entirely due to confinement, as in the MIT bag model, the background scalar field of the NJL model would be irrelevant for nuclear physics since its coupling  to the nucleon would vanish. The reason is very simple as the quarks inside the bag, a bubble of perturbative vacuum, do not feel the presence of the surrounding scalar field. 
Said differently the constituent quarks to which this scalar field couples are in this case absent in  the nucleon. 
 It is likely that the nucleon mass has a mixed origin, in part from chiral symmetry breaking and in part 
from confinement. In this case,  the nucleon mass in the  nuclear medium can feel the presence of the scalar field of the NJL model. At the same time it reacts against this field and it is possible to stabilize nuclear matter. Nuclear  saturation may then result from  a delicate balance between the influence of chiral symmetry breaking and confinement in nucleon structure.
It is clear however that the importance of the role played by the background scalar field in the nuclear binding and saturation 
should not be left to prejudice and beliefs  but to facts which may help  elucidate this role. This is among the purposes of this article.
Some questions to be answered are~: if the scalar field is an actor  in the
nuclear binding and saturation problem, is this role quantitatively compatible with  nuclear phenomenology?  What  informations do we have 
on the role played by confinement? Is it compatible with    
acceptable models of the nucleon structure? The last question  is a motivation for the second part  of this work where we propose  models of the nucleon where  confinement and chiral symmetry breaking contribute roughly equally to the mass. The influence of the spontaneous breaking of chiral symmetry is large enough for the background scalar field to act as a source of  nuclear attraction.  But the confining aspect is sufficient to stabilize nuclear matter.

Our article is organized as follows. The second section is devoted to the Nambu-Jona-Lasinio model. After a brief reminder of the basic properties of the model, we derive from it an effective lagrangian  which is valid for low (space-like) momenta relevant for nuclear physics studies. We first use a sharp momentum cutoff and in a final step for practical calculations we  use a delocalized version. We conclude this section by some general comments concerning in particular the evolution of the nucleon mass and quark condensate when confinement effects at the level of the nucleon structure are incorporated. In the last section, based on a simple quark-diquark string model,  we discuss  how the interplay between chiral symmetry breaking and confinement in the nucleon structure influences nuclear matter binding properties. We also discuss the influence of the modeling of the confinement mechanism inside the nucleon.
\section{The scalar background field from the Nambu-Jona-Lasinio model}

\subsection{The standard NJL model}

We first introduce the NJL model in  the light quark sector whose original  aim is to describe the low mass  mesons:  the pion, the sigma, the rho, the $a_1$ and the omega mesons. The  lagrangian is~:
\begin{eqnarray}
{\cal L}&=& \bar{\psi}\left(i\,\gamma^{\mu}\partial_\mu\,-\,m\right)\,\psi\,+\,\frac{G_1}{2}\,\left[\left(\bar{\psi}\psi\right)^2\,+\
\left(\bar{\psi}\,i\gamma_5\vec\tau\,\psi\right)^2\right]\nonumber\\
& &-\,\frac{G_2}{2}\,\left[\left(\bar{\psi}\,\gamma^\mu\vec\tau\,\psi\right)^2\,+\,
\left(\bar{\psi}\,\gamma^\mu\gamma_5\vec\tau\,\psi\right)^2\,+\,\left(\bar{\psi}\,\gamma^\mu\,\psi\right)^2\right].
\end{eqnarray}
Using path integral technics it can be equivalently  written in a semi-bozonized form~:
\begin{eqnarray}
{\cal L}&=& \bar{\psi}\left[i\,\gamma^{\mu}\partial_\mu\,-\,m\,-\,\Sigma\,-
\,i\,P\,\gamma^5\,-\,\gamma^{\mu}\left(\tilde V_\mu\,+\,\gamma_5\,\tilde A_\mu\right)\right]\psi\nonumber\\
& &-\frac{1}{4\,G_1}\,tr_f\left(\Sigma^2\,+\,P^2\right)\,+\,
\frac{1}{4\,G_2}\,tr_f\left(\tilde{V}^\mu \tilde V_\mu\,+\,\tilde A^\mu \tilde A_\mu\right).
\end{eqnarray}
Here    $\psi$ represents an isodoublet of quark fields, $\Sigma$ is a scalar-isoscalar field, the matrix $P=\vec{\tau}\cdot\vec{P}\equiv \tau_j P_j$ describes  a pseudo-scalar isovector field. The matrix 
$\tilde{V}^\mu=\tilde{\Omega}^\mu\,+\,{\tau}_j\tilde{V}_j^\mu$ contains  an isoscalar ($\tilde\Omega$) and an isovector ($\tilde{V}_j$) vector fields and $\tilde{A}^\mu={\tau}_j\cdot\tilde{A}_j^\mu$ is an isovector axial-vector field. The current quark mass is $m$ and $G_1$ and $G_2$ are two (positive) coupling constants. $\vec P$ is the chiral partner of the $\Sigma$, $\tilde A_j$ is the chiral partner of $\tilde V_j$ and in the limit of vanishing $m$ (chiral limit) this lagrangian is chiral invariant. Coupling to  left (${\cal L}^\mu$) and right 
(${\cal R}^\mu$) electroweak currents is included through the replacement~: 
$$
\tilde V_\mu + \gamma_5\,\tilde A_\mu\to \tilde V_\mu +\gamma_5 \tilde A_\mu\,
+\,{\cal V}_\mu + \gamma_5 {\cal A}_\mu\quad\hbox{with}\quad 
{\cal V}^\mu=\frac{{\cal R}^\mu + {\cal L}^\mu}{2},\quad 
{\cal A}^\mu=\frac{{\cal R}^\mu - {\cal L}^\mu}{2}.
$$
In the mean field approximation the constituent quark mass $M_0$ is solution of the gap equation
\begin{equation}
M_0=m\,+\,4\,N_c\,N_f\,\,G_1\,M_0\,I_1(M_0)\qquad\hbox{with}\qquad I_1=	\int\,\frac{i\,d^4 k}{(2\,\pi)^4}\frac{1}{k^2\,-\,M_0^2}=
\int_0^\Lambda\,\frac{d^3 k}{(2\,\pi)^3}\frac{1}{2\,E_k}\label{GAP}.
\end{equation}
where $N_c=3$ and $N_f=2$ are the number of colors and flavors and $E_k=\sqrt{k^2+M_0^2}$. The quark condensate (per flavor) is $\left\langle \bar q q\right\rangle=-\,4\,N_c\,M_0\,I_1$.
The second form of the $I_1$ integral has been obtained through the introduction of a sharp three-momentum cutoff $\Lambda$.  This sharp non covariant cutoff is first taken for simplicity. We will use later a delocalized version of the NJL which corresponds to a softer cutoff procedure.  The mesons can be generated as collective $q\bar q$ modes either by applying  standard RPA to the original lagrangian or by performing a second order expansion in the fluctuating fields of the bosonized effective action.
We list here some results and for that purpose we introduce the integral~:
\begin{equation}
I(\omega)\equiv	2\,N_c\,N_f\,I_2(\omega)=2\,N_c\,N_f\,\int_0^\Lambda\,\frac{d^3 k}{(2\,\pi)^3}\frac{1}{E_k\,(4\,E_k^2\,-\,\omega^2)}.
\end{equation}
The $q\bar q$ scattering amplitude at zero CM momentum, ${\cal M}^{(\pi)}(\omega)$,  in the pion channel is obtained from the polarization bubble $\tilde\Pi^0_{PS}({\omega})$ in the pseudoscalar channel  incorporating the $\pi-a_1$ mixing: 
\begin{equation}
{\cal M}^{(\pi)}(\omega)=\frac{G_1}{1\,-\,G_1\,\tilde\Pi^0_{PS}({\omega})}\equiv\frac{1/\tilde I(\omega)}{\omega^2\,-\,M^2_\pi(\omega)}	
\qquad\hbox{with}\qquad M^2_\pi(\omega)=\frac{m}{G_1\,M_0\,\tilde I(\omega)}.
\end{equation}
The difference between  $\tilde I(\omega)=I(\omega)/\left[1\,+\,4\,M_0^2\,G_2\,I(\omega)\right]$ and  $I(\omega)$ comes from the mixing effect. From this result we deduce the physical pion mass, the pion-quark coupling constant and the pion decay constant~:
\begin{equation}
m^{2}_{\pi}	=M^{2}_{\pi}(m_{\pi})\quad g^2_{\pi qq}=\frac{r_\pi}{\tilde I(m_\pi)}\qquad 
f^{2}_{\pi}=M^{2}_{0}\,\tilde I(m_\pi)\,r_\pi\qquad
\end{equation}
Here the factor $r_\pi=\left[1\,+\,\frac{m^{2}_{\pi}}{\tilde I(m_\pi)}
\left(\frac{\partial\tilde I}{\partial\omega^2}\right)\right]^{-1}$, which in practice is very close to one, is the residue at the pion pole.
\subsection{Effective theory for low-momentum nuclear physics}

The meson spectrum (scalar and vector mesons) can in principle be  obtained in the previous scheme. This  approach  is notoriously  unsatisfactory due to the lack of confinement: in particular unphysical decay channels of vector mesons in $q \bar q$ pairs may appear but, as discussed in the introduction  we aim to derive an effective low momentum theory ( {\it i.e.} for low space-like momenta relevant in nuclear physics) and not to discuss the on-shell properties of scalar and vector mesons, in particular  their physical masses. Hence our resulting mass parameters for scalar and vector mesons will not be the on-shell masses but simply mass parameters  associated with the inverse of the corresponding correlators taken at zero momentum. As emphasized by Celenza et al \cite{CSWSX95,CWS01} confinement, which is needed to prevent unphysical decays of mesons,  plays a  minor role for the low momentum 
fields (in particular the scalar one) relevant in nuclear physics.  We now  describe the technical steps needed to establish the form of the effective low-momentum lagrangian. 

\paragraph
{Effective lagrangian from NJL model.}

\smallskip\noindent
The aim  is to establish a low momentum lagrangian in the meson sector so as to generate the dynamics of the scalar field. Technically this can be done by integrating out quarks in the Dirac sea using a path integral formalism. The physical meaning is simply a projection of $q\bar q$ vacuum fluctuations onto mesonic degrees of freedom.

In the spirit of our previous works \cite{CEG01,CE05, CE07, MC08} we first go from the cartesian representation, ($\Sigma, P$), to a polar representation, (${\cal S},\pi$),  by making the change of variables~:
\begin{equation}
m\,+\,\Sigma	\,+\,i\,P={\cal S}\,U\quad\hbox{with}\quad U\equiv\xi^2=e^{i\,\vec\tau\cdot\vec\pi}.
\end{equation}
It is convenient to  introduce a new quark field $q$ defined by~:
\begin{equation}
q=\xi_5\,\psi,\quad\bar{q}=\bar{\psi}\,\xi_5\qquad\hbox{with}\qquad	\xi_5=e^{i\,\vec\tau\cdot\vec\pi\gamma_5/2}
\end{equation}
and new vector and axial vector fields according to~:
\begin{eqnarray}
V^\mu &=&\xi\,\frac{\tilde{V^\mu}\,+\,\tilde{A^\mu}}{2}\,\xi^\dagger\,+\,\xi^\dagger\,\frac{\tilde{V^\mu}\,-\,\tilde{A^\mu}}{2}\,\xi\,-\,
{\cal V}^\mu_c\,+	\,{\cal V}_\xi^\mu\nonumber\\
A^\mu &=&\xi\,\frac{\tilde{V^\mu}\,+\,\tilde{A^\mu}}{2}\,\xi^\dagger\,-\,\xi^\dagger\,\frac{\tilde{V^\mu}\,-\,\tilde{A^\mu}}{2}\,\xi\,-\,
{\cal A}^\mu_c\,+	\,{\cal A}_\xi^\mu
\end{eqnarray}
with
\begin{eqnarray}
{\cal V}^\mu_c &=& \frac{i}{2}\left(\xi\partial^\mu\xi^\dagger\,+\,\xi^\dagger\partial^\mu\xi\right),\qquad
{\cal A}^\mu_c=\frac{i}{2}\left(\xi\partial^\mu\xi^\dagger\,-\,\xi^\dagger\partial^\mu\xi\right)\nonumber\\
{\cal V}^\mu_\xi &=& \frac{1}{2}\left(\xi{\cal R}^\mu\xi^\dagger\,+\,\xi^\dagger{\cal L}^\mu\xi\right),\qquad
{\cal A}^\mu_\xi = \frac{1}{2}\left(\xi{\cal R}^\mu\xi^\dagger\,-\,\xi^\dagger{\cal L}^\mu\xi\right).
\end{eqnarray}
With these new fields the semi-bozonized lagrangian takes the form~:
\begin{eqnarray}
{\cal L}&=& \bar{q}\left[i\,\gamma^{\mu}\partial_\mu\,-\,{\cal S}\,-\,\gamma^{\mu}\left( V_\mu\,+\,\gamma_5\, A_\mu\right)\right]q
\,-\,\frac{1}{4\,G_1}\,tr_f\left({\cal S}^2\,-\,m\,{\cal S}\,(U+U^\dagger)\right)\nonumber\\
& & \,+\,\frac{1}{4\,G_2}\,tr_f\left(({V}^\mu +{\cal V}^\mu_c -{\cal V}^\mu_\xi)^2\,+\,
({A}^\mu +{\cal A}^\mu_c -{\cal A}^\mu_\xi)^2\right).
\end{eqnarray}
The next step is to integrate out the quarks in the Dirac sea. In that way the kinetic energy term of the mesons fields will be dynamically generated from the quarks loops, {\it i.e}, from quantum fluctuations. For convenience we go from Minkovski space to Euclidean space. Using standard transformation rules   the corresponding Euclidean lagrangian  is~: 
\begin{equation}
L_E=\bar q\,D\,q\quad\hbox{with}\quad D=i\,\gamma^E_\mu\cdot \Pi_\mu\,+\,{\cal S},\quad	\Pi^\mu=P_\mu-\,\Gamma_\mu\equiv
-i\partial_\mu\,-\,(V_\mu+\gamma_5 A_\mu).
\end{equation}
The Euclidean partition function is expressed in term of the fermion determinant according to~: 
\begin{equation}
Z=e^{-S_F}= 	\int\,dq\,d\bar{q}\,e^{-d^4x\,\bar q\,D\,q}.
\end{equation}
Ignoring its imaginary part, the effective action can be written as~: 
\begin{eqnarray}
S_F&=&- Tr ln D=-\frac{1}{2}	Tr ln (D D^\dagger)\nonumber\\
&=&-\frac{1}{2}	Tr ln\left(\Pi^2_,+i\alpha_{\mu\nu}W_{\mu\nu}+{\cal S}^2-\gamma_{\mu}{\cal D}_{\mu}{\cal S}\right)
\end{eqnarray}
with
\begin{eqnarray}
& &\Pi_{\mu}=P_\mu	- \Gamma_{\mu}\equiv -i\partial_{\mu}-\left(V_{\mu}+\gamma_5 A_{\mu}\right),\qquad
{\cal D}_{\mu}{\cal S}=\partial_{\mu}{\cal S}+2i\gamma_5A_{\mu}\nonumber\\
& &W_{\mu\nu}=\partial_{\mu}\Gamma_{\nu}-\partial_{\nu}\Gamma_{\mu}- i\left[\Gamma_{\mu},\Gamma_{\nu}\right]
\end{eqnarray}
We perform a derivative expansion valid at low momentum of the fermion determinant to second order in the derivatives. The difficulty lies in the fact that we do not make an expansion  around a constant (vacuum expectation value of the scalar field)  but we want to have a formal expansion with the scalar objet ${\cal S}$ keeping its field status, so as to include 
its possible modification in the nuclear environment. For that purpose we use the  elegant method proposed by Chan \cite{CHAN86}. The starting point is  the  following trick which uses translational invariance in momentum space:
\begin{equation}
Tr \left[A\left(\Pi_{\mu},G(X)\right)\right] =	Tr \left[e^{i k\cdot X} A\left(\Pi_{\mu},G(X)\right)e^{-i k\cdot X}\right]=
Tr A\left(\Pi_{\mu}+ k_\mu ,G(X)\right)
\end{equation}
where $A$ represents any operator depending of the position operator $X$ and of the generalized momentum $\Pi$. Hence the arbitrary four-momentum, $k$,  can be averaged~:
\begin{equation}
Tr \left[A\left(\Pi_{\mu},G(X)\right)\right]= 
\frac{1}{\delta^{(4)}(0)}\int\frac{d^4 k}{(2\pi)^4}Tr A\left(\Pi_{\mu}+ k_\mu ,G(X)\right).
\end{equation}
 It follows that the quark determinant can be calculated as~:
\begin{eqnarray}
S_F&=&-\frac{1}{2}\,\frac{1}{\delta^{(4)}(0)}\int\frac{ d^4 k}{(2\pi)^4} Tr\left[ln(G^{-1}) + ln\left(1 + G \Pi^2 +G (2 k\cdot\Pi+a\right)\right]\nonumber\\
& &\hbox{with}\qquad G=\left(k^2 + {\cal S}^2\right)^{-1},\qquad a= i\alpha_{\mu\nu}W_{\mu\nu} - \gamma_{\mu}{\cal D}_{\mu}{\cal S}.\label{SEFF}
\end{eqnarray}
As pointed out by Chan, the introduction of the momentum integration  does  not disturb the full trace operation and offers the  freedom needed for manipulations, such as cyclic permutations of the operators 
or integrations by part under the condition that there is an implicit regularization scheme. The essential point is that the final form  for the action  is entirely expressible in terms of covariant derivatives $\left[\Pi_{\mu},{\cal S}\right]$ and $\left[\Pi_{\mu},\Pi_{\nu}\right]=W_{\mu\nu}$, as it should. Once this is done the explicit trace over $\left|x\right\rangle$ states can be performed producing the ${\delta^{(4)}(0)}$ compensating the one coming from the average procedure. 
To perform a second order derivative expansion corresponds in practice to make a fourth order expansion in $\Pi_{\mu}$ and second order in $a$. The result is~:
\begin{eqnarray}
S_F=-\frac{1}{2}\,\int\frac{ d^4 k\, d^4x}{(2\pi)^4}& &tr_{DFC}\bigg(ln(G^{-1})\,-\,\frac{1}{2}G^2\,a^2\nonumber\\
& &+\,k^2\,G^2\,{\cal S}^2\,\partial_{\mu}{\cal S}\partial_{\mu}{\cal S}
\,+\,\frac{k^2 k^2}{12}\,W_{\mu\nu}W_{\mu\nu}\bigg)\label{SGENERAL}
\end{eqnarray}
where the trace operation acts in Dirac, flavor and color spaces.
The resulting effective action is reducible to a local lagrangian. Coming back to Minkovski space but keeping the momentum $k$  explicitly in Euclidean space this local Lagrangian, ${\cal L}_{mes}$,  has the form~:
\begin{eqnarray}
{\cal L}_{mes}&=& \frac{1}{2}\,2 N_c N_f\,I_{2S}\,({\cal S})\,\partial^{\mu}{\cal S}\,\partial_{\mu}{\cal S}\,- \,W({\cal S})
\nonumber\\
& & +\,\frac{m\,{\cal S}}{4\,G_1}\,tr_f(U\, +\, U^\dagger\,-\,2)\,+\,\frac{1}{2}\,2 N_c N_f\,I_{2}({\cal S})\, 4\vec{A}^\mu\cdot \vec{A}_\mu\,{\cal S}^2\nonumber\\
& &\frac{1}{4\,G_2}\,tr_f\big(({V}^\mu +{\cal V}^\mu_c -{\cal V}^\mu_\xi)^2\,+\,
({A}^\mu +{\cal A}^\mu_c -{\cal A}^\mu_\xi)^2\big)\nonumber\\
& &-\,\frac{1}{6}\,2 N_c N_f\,I_{2V}({\cal S})\,\left(\Omega^{\mu\nu}\Omega_{\mu\nu}\,+\,\vec{V}^{\mu\nu}\cdot \vec{V}_{\mu\nu}\,+\,\vec{A}^{\mu\nu}\cdot \vec{A}_{\mu\nu}\right).\label{EFFLAG}
\end{eqnarray}
 The chiral effective potential, $W({\cal S})$, is~:
\begin{equation}
W({\cal S})=-	2 N_c N_f\,I_{0}\,({\cal S})\,+\,\frac{\left({\cal S}\,-\,m\right)^2}{2\,G_1}.
\end{equation}
the quantity, $-I_{0}\,({\cal S})$, represents the vacuum energy density per degrees of freedom associated with the Dirac sea
\begin{equation}
I_{0}\,({\cal S})=	\int\,\frac{i\,d^4 k_E}{(2\,\pi)^4}\, ln\left(k_E^2\,+\,{\cal S}^2\right)=
\int_0^\Lambda\,\frac{d^3 k}{(2\,\pi)^3}\,E_k({\cal S})\qquad	E_k\equiv E_k({\cal S})=\sqrt{k^2\,+\,{\cal S}^2}
\end{equation}
where the second form corresponds to the sharp non covariant cutoff.  The integral $I_{2}\,({\cal S})$ is the usual NJL loop integral~:
\begin{equation}
I_{2}\,({\cal S})=	\int\,\frac{i\,d^4 k_E}{(2\,\pi)^4}\frac{1}{\left(k_E^2\,+\,{\cal S}^2\right)^2}=
\int_0^\Lambda\,\frac{d^3 k}{(2\,\pi)^3}\frac{1}{4\,E^3_k({\cal S})}.
\end{equation}
The integrals $I_{2S,V}\,({\cal S})$ entering the scalar and vector kinetic energy terms are~:
\begin{eqnarray}
& & I_{2S}=I_2 - 2 {\cal S}^2 I_3 +2  {\cal S}^4 I_4,\qquad         I_{2V}=I_2 + {\cal S}^2 I_3 -\frac{1}{2}  {\cal S}^4 I_4\nonumber\\
& & I_{3}\,({\cal S})=	\int\,\frac{i\,d^4 k_E}{(2\,\pi)^4}\frac{1}{\left(k_E^2\,+\,{\cal S}^2\right)^3}=
\int_0^\Lambda\,\frac{d^3 k}{(2\,\pi)^3}\frac{3}{16\,E^5_k({\cal S})}\nonumber\\
& & I_{4}\,({\cal S})=	\int\,\frac{i\,d^4 k_E}{(2\,\pi)^4}\frac{1}{\left(k_E^2\,+\,{\cal S}^2\right)^4}=
\int_0^\Lambda\,\frac{d^3 k}{(2\,\pi)^3}\frac{5}{96\,E^7_k({\cal S})}.
\end{eqnarray}
This expression of the lagrangian shows that our objective of eliminating the vacuum $q\bar q$ fluctuations in terms of ``observable'' background fields is realized. 
Contrary to usual methods   the integrals $I_{2S,V}$ appearing in the scalar and vector kinetic energy terms differ from  the usual $I_2$ integral. However  the difference between $I_{2S,V}$ and $I_2$  has no influence on nuclear matter calculation, at least in the Hartree approximation, since the derivative terms play no role. It is also important to notice that  the derivation  has been done assuming implicitly a covariant regularization procedure and may not be strictly valid for the non covariant cutoff which will be used in practice. 

The meson-like lagrangian written above in eq. (\ref{EFFLAG}) has been obtained by integrating out the fluctuating quark fields (quarks in the Dirac sea). It remains  to enlarge  this lagrangian  to the ``classical'' quark fields corresponding to the valence quark sector. Formally this can be done by introducing quark source terms and splitting the quark field  into a classical part $Q$ and a fluctuating part. The integration of the fluctuation part produces  the above quark determinant and the lagrangian ${\cal L}_{mes}$.   After a Legendre transformation, we arrive at the valence quark effective lagrangian which simply reads~:
\begin{equation}
{\cal L}_{val}= \bar{Q}\left[i\,\gamma^{\mu}\partial_\mu\,-\,{\cal S}\,-\,\gamma^{\mu}\left( V_\mu\,+\,\gamma_5\, A_\mu\right)\right]Q.
\end{equation}
 
Remind that ${\cal S}$ is still a field and its vacuum expectation value (corresponding to the 
minimum of $W({\cal S})$) is the vacuum constituent quark mass,  $M_0$, solution of the gap equation (\ref{GAP}). In nuclear matter its expectation value, $\bar {\cal S}$, is the solution of an in-medium gap equation modified by the presence of the nucleonic scalar density. It coincides with the in-medium modified constituent quark mass $M$ (see next subsection). Its fluctuation  enters the scalar exchange Fock term according to  the treatment given in \cite{MC08}. In the following we will replace in eq. (\ref{EFFLAG})  the scalar field by  its expectation value in all derivatives terms and in the extra mass term, $I_{2}({\cal S})\, 4\vec{A}^\mu\cdot \vec{A}_\mu\,{\cal S}^2$, for the axial vector meson. In order to prepare the identification of canonical modes, we rewrite  (omitting electroweak fields) the mesonic lagrangian as
\begin{eqnarray}
{\cal L}_{mes}&=& \frac{1}{2}\,2 N_c N_f\,I_{2S}(\bar{\cal S})\,\partial^{\mu}{\cal S}\,\partial_{\mu}{\cal S}\,- \,W({\cal S})\nonumber\\
& & +\,\frac{m\,{\cal S}}{4\,G_1}\,tr_f(U\, +\, U^\dagger\,-\,2)\,+\,
\tilde I(\bar{\cal S})\,\bar{\cal S}^2\,tr_f\left({\cal A}^\mu_c\cdot{\cal A}^c_\mu\right)
\nonumber\\
& &
+\,\frac{1}{4\,G_2}\,\left(1\,+\,4\,G_2\,I(\bar{\cal S})\,\bar{\cal S}^2\right)\,tr_f\left({A}^\mu \,+\,\frac{{\cal A}^\mu_c}
{1\,+\,4\,G_2\,\bar{\cal S}^2\,I(\bar{\cal S})}\right)^2\nonumber\\
& &+\,\frac{1}{4\,G_2}\,tr_f\left({V}^\mu +{\cal V}^\mu_c \right)^2\,-\,\frac{1}{6}\,2 N_c N_f\,I_{2V}(\bar{\cal S})\,\left(\Omega^{\mu\nu}\Omega_{\mu\nu}\,+\,\vec{V}^{\mu\nu}\cdot \vec{V}_{\mu\nu}\,+\,\vec{A}^{\mu\nu}\cdot \vec{A}_{\mu\nu}\right)
\end{eqnarray}
where we have introduced the quantities~:
$$I(\bar{\cal S})=2 N_c N_f\, I_{2}(\bar{\cal S}),\qquad
\tilde I(\bar{\cal S})\equiv 2 N_c N_f\,\tilde I_{2}(\bar{\cal S})=\frac{I(\bar{\cal S})}{1\,+\,4\,G_2\,\bar{\cal S}^2\,I(\bar{\cal S})}.$$
We now redefine the axial-vector meson field in order to eliminate the $\pi-a_1$ mixing. For this purpose we introduce the canonical axial vector field, $a_\mu$,   defined according to~:
\begin{equation}
	{A}^\mu \,+\,\frac{{\cal A}^\mu_c}
{1\,+\,4\,G_2\,\bar{\cal S}^2\,I(\bar{\cal S})}=g_V\,a^\mu\qquad \hbox{with}\qquad g^{2}_{V}=\frac{3/2}{2 N_c N_f\,I_{2V}(\bar{\cal S})}
\end{equation}
where $g_V$ is the quark-vector coupling constant.
Similarly the canonical vector ($\omega_\mu, v_\mu$)  and scalar ($S$),  fields are defined as~:
\begin{eqnarray}
& &	\Omega^{\mu}=g_V\,\omega^{\mu},\qquad\qquad {V}^{\mu}=g_V\,{v}^{\mu}\nonumber\\
& & {\cal S}=g_{0S}\, S \qquad \hbox{with}\qquad g^2_{0S}=\frac{1}{2 N_c N_f\,I_{2S}(M_0)}.\label{QUARKSCALAR}
\end{eqnarray}
Here the quark-scalar coupling constant, $g_{0S}$, is defined at the vacuum point. 
Omitting the $\rho\pi\pi$ coupling terms,  the low-momentum effective lagrangian takes the form:
\begin{eqnarray}
{\cal L}_{mes}&=&\frac{1}{2}\,\frac{I_{2S}(\bar{\cal S})}{I_{2S}(M_0)}\partial^{\mu} S\partial{\mu}S\,+\,W	({\cal S}=g_{0S} S)\nonumber\\
& &+\frac{m\,{\cal S}}{4\,G_1}\,tr_f(U\, +\, U^\dagger\,-\,2)\,+\,
\tilde I(\bar{\cal S})\,\bar{\cal S}^2\,tr_f\left({\cal A}^\mu_c\cdot{\cal A}^c_\mu\right)
\nonumber\\
& &+ \frac{1}{2}M^{2}_{V}\,\left(\omega^{\mu}\omega_{\mu}\,+\,\vec{v}^{\mu}\cdot\vec{v}_{\mu}\right)\,+\,
\frac{1}{2}M^{2}_{A}\,\left(\vec{a}^{\mu}\cdot\vec{a}_{\mu}\right)\nonumber\\
& &-\,\frac{1}{4}\,\left(\omega^{\mu\nu}\omega_{\mu\nu}\,+\,\vec{v}^{\mu\nu}\cdot \vec{v}_{\mu\nu}\,+\,\vec{a}^{\mu\nu}\cdot \vec{a}_{\mu\nu}\right).\label{LFINAL}
\end{eqnarray}
The vector and axial-vector low momentum mass parameters are given by~~:
\begin{equation}
 M^2_V=\frac{g^{2}_{V}}{G_2},\qquad\qquad\frac{M^2_A}{M^2_V}=1\,+\,4\,G_2\,I(\bar{\cal S})\,\bar{\cal S}^2.
\end{equation}
One  defines the canonical pion field, $\Phi\equiv\vec\tau\cdot\vec\Phi$, through $U=exp(i\Phi/F)$ where the constant $F$ is given by  $F^2=\tilde I(M_0)\,M^{2}_{0}$.
Coming back to the previous form of the lagrangian (eq. \ref{EFFLAG}), a direct inspection of the coupling of $\partial^\mu\Phi$ to the axial weak current allows to identify the pion decay constant $F_\pi$ with the parameter $F$. The vacuum pion mass parameter  is finally obtained as~:
\begin{equation}
M^{2}_{\pi}=\frac{m\,M_0}{G_1\,F^2_\pi}.
\end{equation}
In nuclear matter the explicit ${\cal S}$ factor in front of the pion mass term  renormalizes the pion mass. As in ref. \cite{CE07} we do not consider this effect since it is almost completely compensated by other contributions and the pion mass is expected to remain stable in nuclear matter.
As for the (canonical) vacuum scalar mass, it  is~:
\begin{equation}
M^2_\sigma=	g^2_{0S}\,\left(\frac{\partial^2 W}{\partial {\cal S}^2}\right)_{{\cal S}=M_0}=\frac{I_{2}(M_0)}{I_{2S}(M_0)}\,
\left(4\,M^{2}_{0}\,+\,\left(\frac{M^2_V}{M^2_A}\right)_{vac}M^{2}_{\pi}\right).
\end{equation}
We stress again that the quantities $M_V, M_A, M_\pi, F_\pi$ are not on-shell constants but  low momentum effective lagrangian parameters, ({\it i.e.}, taken at zero momentum).  In practice however $M_\pi$ and $F_\pi$ differ little from the physical pion mass and pion decay constant. On the contrary the vector and axial vector mass parameters have {\it a priori}  no reason to coincide with the physical omega, rho and $a_1$ meson masses. We now develop an alternative approach which relaxes the sharp cutoff procedure.

\paragraph{Delocalized NJL model.} 

\smallskip\noindent
As seen before  when a vector interaction term is added the $\pi-a_1$ mixing has the effect of decreasing  the pion decay constant and it   is not  easy  with the sharp cutoff to reach a  sufficiently large value for $F_\pi$. This is one motivation to adopt  for practical phenomenological calculations  another version of the NJL model with a smooth cutoff function. As discussed below there is also physical motivation for such a smooth regularization associated with non localities. 

The non local version of the NJL model is obtained, for any channel,  with the replacement~:
\begin{equation}
\left(\bar{\psi}\Gamma_j\psi\right)(x)\,\to\,J_j(x)=\int d^4 x_1\,d^4 x_2\,	F_c(x_1-x)\,F_c(x-x_2)\,\bar\psi(x_1)\Gamma_j\psi(x_2).\label{DELOC}
\end{equation}
The presence of the form factor  automatically provides a regularization procedure.  Moreover such a  delocalized lagrangian possesses a physical basis in terms of quark-instanton interaction \cite{RWB04}.
We define the Fourier  transform, $f(p)$,  of the form factor appearing in the delocalized currents $J_j(x)$,
\begin{equation}
F_c(x)=\int \frac{d^4 p}{(2\pi)^4}	\,e^{-ip\cdot x}\, f(p),
\end{equation}
with $f(0)=1$ in such a way that $F_c$ satisfies the normalization condition $\int d^4x \,F_c(x)=1$.
As can be checked  this procedure maintains the chiral invariance of the interaction. For actual calculations we will take the non covariant version of the non local NJL model~:
\begin{equation}
F_c(x)=\delta(t)\,F(\vec{r}),\qquad f(p)\equiv f(\vec{p})=\int d^3 r\,e^{-i \vec{p}\cdot\vec{r}}\,F(\vec{r}).
\end{equation}
One  practical consequence is that the interaction when written in momentum space is modified according to~:
\begin{equation}
	G_j \delta^{(3)}(\vec{p}_1 + \vec{p}_2 - \vec{p}_3 - \vec{p}_4)\quad\to\quad
G_j \delta^{(3)}(\vec{p}_1 + \vec{p}_2 - \vec{p}_3 - \vec{p}_4)\,f(\vec{p}_1)f(\vec{p}_2)f(\vec{p}_3)f(\vec{p}_4).
\end{equation}
As it is always implicitly done  in NJL (with various cutoff prescriptions), we also apply the delocalization procedure to the current quark mass term:
$$m\,\bar{\psi}(x)\psi(x)\quad\to\quad
m\, \int d^4 x_1\,d^4 x_2	\,F_c(x_1-x)\,F_c(x-x_2)\,\bar\psi(x_1)\psi(x_2).$$
 It can be checked  that the QCD realization of explicit chiral symmetry breaking is not affected in the sense that the operator identity,
$\left[Q_i\left[Q_j,H\right]\right]=H_{\chi SB}$, is still realized and consequently the GOR relation also holds. In the following we  choose a gaussian  for the form factor
\begin{equation}
f(\vec{p})=e^{\frac{-p^2}{\,2\,\Lambda^2}}	
\end{equation}
where $\Lambda$ (possibly related to the inverse of the instanton size) is the cutoff parameter of the order of
 $1$ GeV. One advantage of the non local version is the smooth momentum dependence of the constituent quark
 mass, in agreement with lattice calculation. Indeed the gap equation in vacuum writes~:
\begin{eqnarray}
& & M(\vec{p})= M_0\,f^2(\vec{p}), \qquad\hbox{with}\qquad M_0=m\,-\,2\,G_1\,\left\langle \bar q q\right\rangle	\qquad\nonumber\\
& & \left\langle \bar q q\right\rangle =-N_c N_f\,\int\frac{d^3 k}{(2\pi)^3}\,f^2(\vec{k})\,\frac{M(k)}{E_k}\qquad
E_k=\sqrt{k^2+M^2(k)}.
\end{eqnarray}
The delocalized  version of the semi-bozonized form of eq. 2 writes~:
\begin{eqnarray}
{\cal L}(x)&=& \bar{\psi}(x)i\,\gamma^{\mu}\partial_\mu\,\psi(x)\nonumber\\
&- & \int d^4x_1 d^4 x_2\,\bar{\psi}(x_1)\,F_c (x_1-x)\left(\Sigma\,+
\,i\,P\,\gamma^5\,+\,\gamma^{\mu}(\tilde V_\mu\,+\,\gamma_5\,\tilde A_\mu)\right)(x)\,F_c (x-x_2)\,\psi(x_2)\nonumber\\
&- &\frac{1}{4\,G_1}\,tr_f\left((\Sigma - m)^2\,+\,P^2\right)(x)\,+\,
\frac{1}{4\,G_2}\,tr_f\left(\tilde{V}^\mu \tilde V_\mu\,+\,\tilde A^\mu \tilde A_\mu\right)(x)
\end{eqnarray}
In the presence of the form factor we found more convenient  to perform the quark integration with the
 original quark field. Going again in Euclidean space the Dirac operator, $D^l$,  is defined according to
\begin{eqnarray}
& & D^l	=i\gamma^{E}_{\mu}\cdot{\Pi}^{l}_{\mu} + {W}^{l},\qquad {\Pi}^{l}_{\mu}=P_{\mu}-{\Gamma}^{l}_{\mu}\nonumber\\
& & {\Gamma}^{l}_{\mu}=\hat{F}(P)\left(\tilde{V}_{\mu} + \gamma_5 \tilde{A}_{\mu}\right)\hat{F}(P)\nonumber\\
& & {W}^{l}=\hat{F}(P) W \hat{F}(P),\qquad W=\Sigma + i \gamma_5 P\equiv {\cal S}\, U_5=\equiv {\cal S}\xi^{2}_{5}
\end{eqnarray}
where $\hat{F}(P)$ is an operator diagonal in momentum space whose eigenvalues coincide with  the form 
factor $f(p)$. We can see that the effect of non local 
coupling is to transform all the field operators $O(X)$ into $O^l(X)=\hat{F}(P) O(X) \hat{F}(P)$. Inclusion
  of electroweak coupling is done by making the replacement of the type:~$\hat{F}(P) \tilde{A}_\mu \hat{F}(P)\to  \hat{F}(P) \tilde{A}_\mu \hat{F}(P) +{\cal A}_\mu$.

\noindent
The euclidean effective action is 
	\begin{eqnarray}
S_F&=&- Tr ln D^l=-\frac{1}{2}	Tr ln (D^l D^{l \dagger})\nonumber\\
&=&-\frac{1}{2}	Tr ln\left(\Pi^{l2}\,+i\alpha_{\mu\nu}W^l_{\mu\nu}+{W}^{l}{W}^{l\dagger }-\gamma_{\mu}{\cal D}^l_{\mu} W^{l}\right)
\end{eqnarray}
with 
\begin{eqnarray}
& &{\cal D}^l_{\mu} W^{l}=\partial_{\mu}\Sigma^l - \left\{\tilde A^{l}_{\mu},P^l\right\} 
+ i\gamma^5\left(\partial_{\mu}P^l - \left[\tilde V^{l}_{\mu}, P^l\right] +   \left\{\tilde A^{l}_{\mu},\Sigma^l\right\}\right)\nonumber\\
& &W^l_{\mu\nu}=\partial_{\mu}\Gamma^l_{\nu}-\partial_{\nu}\Gamma^l_{\mu}- i\left[\Gamma^l_{\mu},\Gamma^l_{\nu}\right].
\end{eqnarray}
For  the derivative expansion we again use the momentum averaging trick. It involves  $\hat F(k+P)$ terms and consequently term with  derivatives of $f(p)$. Here we take the prescription of neglecting them. Hence within this approximation all the fields such as  ${\cal S}$ will be multiplied by the number $f^2(k)$ in the various momentum integrals. The calculation of the quark determinant is formally identical to the previous case~:
\begin{eqnarray}
S_F&=&-\frac{1}{2}\int\frac{d^4x d^4 k}{(2\pi)^4} tr\left[ln(G^{-1}) + ln\left(1 + G \Pi^2 +G (2 k\cdot\Pi+a_l\right)\right]\nonumber\\
& &\hbox{with}\qquad G=\left(k^2 + f^2(k){\cal S}^2\right)^{-1},\qquad a_l= i\alpha_{\mu\nu}W_{\mu\nu} - \gamma_{\mu}{\cal D}^l_{\mu} W^{l\dagger}.
\end{eqnarray}
Coming back to Minkowski space we obtain a new effective Lagrangian. We omit again in its expression $\rho\pi\pi$ terms  and terms involving couplings of pion and axial-vector fields to the derivatives of the scalar field~:
\begin{eqnarray}
{\cal L}^{l}_{mes}&=& \frac{1}{2}\,2 N_c N_f\,I^l_{2S}({\cal S})\,\partial^{\mu}{\cal S}\,\partial_{\mu}{\cal S}\,- \,W^l({\cal S})
+\,\frac{m\,{\cal S}}{4\,G_1}\,tr_f(U\, +\, U^\dagger\,-\,2)\nonumber\\
& &+\,\frac{1}{2}\,2 N_c \,tr_f\left(I^l_{2}({\cal S})\,{\cal S}^2\,\partial^{\mu}{U}\,\partial_{\mu}U^\dagger
+\,\,4I^l_{26}({\cal S})\,\partial^{\mu}P \tilde A_\mu \Sigma
+\,4I^l_{28}({\cal S})\,\tilde A^\mu \tilde A_\mu\Sigma^2\right)\nonumber\\
& &+\,\frac{1}{4\,G_2}\,tr_f(\tilde{V}^\mu \tilde V_\mu\,+\,\tilde{A}^\mu \tilde A_\mu)
-\,\frac{1}{6}\,2 N_c N_f\,I_{2V}({\cal S})\,\left(\tilde\Omega^{\mu\nu}\Omega_{\mu\nu}\,+\,\vec{\tilde V}^{\mu\nu}\cdot \vec{\tilde V}_{\mu\nu}\,+\,\vec{\tilde A}^{\mu\nu}\cdot \vec{\tilde A}_{\mu\nu}\right).\nonumber\\ \label{LAGNEW}
& &
\end{eqnarray}
The new chiral effective potential, $W^l({\cal S})$, is
\begin{equation}
W^l({\cal S})=-	2 N_c N_f\,I^l_{0}\,({\cal S})\,+\,\frac{\left({\cal S}\,-\,m\right)^2}{2\,G_1}.\label{HAT1}
\end{equation}
The quantity, $-I^l_{0}({\cal S})$, represents the vacuum energy density associated with the Dirac sea~:
\begin{equation}
I^{l}_{0}=	\int\,\frac{d^4 k_E}{(2\,\pi)^4}\, ln\left(k_E^2\,+\,f^4(k){\cal S}^2\right)=
\int_0^\infty\,\frac{d^3 k}{(2\,\pi)^3}\,E_k({\cal S})\qquad	E_k\equiv E_k({\cal S})=\sqrt{k^2\,+\,f^4(k){\cal S}^2}.
\label{HAT2}
\end{equation}
where the second form corresponds to a non covariant prescription.  The various other integrals are~:
\begin{eqnarray}
& & I^l_{2S}=I^l_2 - 2 {\cal S}^2 I^l_3 +2  {\cal S}^4 I^l_4,\qquad   I^l_{2V}=I^l_2 + {\cal S}^2 I^l_3 -\frac{1}{2}  {\cal S}^4 I^l_4\nonumber\\
& & I^l_{2}=	\int\,\frac{d^4 k_E}{(2\,\pi)^4}\frac{f^4(k)}{\left(k_E^2\,+\,f^4(k){\cal S}^2\right)^2}=
\int_0^\infty\,\frac{d^3 k}{(2\,\pi)^3}\frac{f^4(k)}{4\,E^3_k({\cal S})}\nonumber\\
& & I^l_{3}=	\int\,\frac{d^4 k_E}{(2\,\pi)^4}\frac{f^4(k)}{\left(k_E^2\,+\,f^4(k){\cal S}^2\right)^3}=
\int_0^\infty\,\frac{d^3 k}{(2\,\pi)^3}\frac{3 f^4(k)}{16\,E^5_k({\cal S})}\nonumber\\
& & I^l_{4}=	\int\,\frac{d^4 k_E}{(2\,\pi)^4}\frac{f^4(k)}{\left(k_E^2\,+\,f^4(k){\cal S}^2\right)^4}=
\int_0^\infty\,\frac{d^3 k}{(2\,\pi)^3}\frac{5 f^4(k)}{96\,E^7_k({\cal S})}\nonumber\\
& & I^l_{26}=	\int\,\frac{d^4 k_E}{(2\,\pi)^4}\frac{f^6(k)}{\left(k_E^2\,+\,f^4(k){\cal S}^2\right)^2}=
\int_0^\infty\,\frac{d^3 k}{(2\,\pi)^3}\frac{f^6(k)}{4\,E^3_k({\cal S})}\nonumber\\
& & I^l_{28}=	\int\,\frac{d^4 k_E}{(2\,\pi)^4}\frac{f^8(k)}{\left(k_E^2\,+\,f^4(k){\cal S}^2\right)^2}=
\int_0^\infty\,\frac{d^3 k}{(2\,\pi)^3}\frac{f^8(k)}{4\,E^3_k({\cal S})}.
\end{eqnarray}
Again the pion-axial-vector mixing has to be eliminated. Keeping only the relevant terms for nuclear physics
 purpose we obtain the following effective lagrangian
\begin{eqnarray}
{\cal L}^{l}_{mes}&=&\frac{1}{2}\,\frac{I^l_{2S}(\bar{\cal S})}{I^l_{2S}(M_0)}\partial^{\mu} S\partial{\mu}S\,-\,W	({\cal S}=g_{0S} S)\nonumber\\
& &+\frac{1}{4}\,F^2\,M_\pi^2\,\frac{\bar{\cal S}}{M_0}\,tr_f(U\, +\, U^\dagger\,-\,2)\,+\,
\frac{1}{2 F^2}\tilde I(\bar{\cal S})\,\bar{\cal S}^2\,\partial^{\mu}\vec{\Phi}\partial^{\mu}\vec{\Phi}
\nonumber\\
& &+ \frac{1}{2}M^{2}_{V}\,\left(\omega^{\mu}\omega_{\mu}\,+\,\vec{v}^{\mu}\cdot\vec{v}_{\mu}\right)\,+\,
\frac{1}{2}M^{2}_{A}\,\left(\vec{a}^{\mu}\cdot\vec{a}_{\mu}\right)\nonumber\\
& &-\,\frac{1}{4}\,\left(\omega^{\mu\nu}\omega_{\mu\nu}\,+\,\vec{v}^{\mu\nu}\cdot \vec{v}_{\mu\nu}\,+\,\vec{a}^{\mu\nu}\cdot \vec{a}_{\mu\nu}\right)
\end{eqnarray}
where we have introduced the quantities
\begin{eqnarray}
& & I^l(\bar{\cal S})=2 N_c N_f\, I^l_{2}(\bar{\cal S}),
\qquad \tilde I^l(\bar{\cal S})=2 N_c N_f\,\tilde I^l_{2}(\bar{\cal S})\nonumber\\
& & \tilde I^l(\bar{\cal S})\equiv 2 N_c N_f\,\tilde I^l_{2}(\bar{\cal S})={I^l(\bar{\cal S})-\frac{4 G_2 \bar{\cal S}^2\,I^{l2}_6(\bar{\cal S})}{1\,+\,4\,G_2\,\bar{\cal S}^2\,I^{l}_{8}(\bar{\cal S})},\qquad F^2=\tilde I^l(M_0)\,M^{2}_{0}}.
\end{eqnarray}
The constant $F$ given above defines the canonical pion field $\Phi$ through $U=exp(\Phi/F)$. The canonical 
vector and axial vector fields are defined according to
\begin{equation}
	\tilde\Omega^{\mu}=g_V\,\omega^{\mu}\quad \tilde{V}^{\mu}=g_V\,{v}^{\mu}\quad \tilde{A}^\mu \,+\,\frac{2\,G_2\,\bar{\cal S}^2\,I^{l}_{6}(\bar{\cal S})}
{1\,+\,4\,G_2\,\bar{\cal S}^2\,I^{l}_{8}(\bar{\cal S})}\partial^{\mu}\Phi=g_V\,a^\mu
\end{equation}
with 
\begin{equation}
g^{2}_{V}=\frac{3/2}{2 N_c N_f\,I^l_{2V}(\bar{\cal S})}, \qquad M^2_V=\frac{g^{2}_{V}}{G_2}, \qquad\frac{M^2_A}{M^2_V}=1\,+\,4\,G_2\,I^{l}_{8}(\bar{\cal S})\,\bar{\cal S}^2.
\end{equation}
Coming back to the previous form of the lagrangian (eq. \ref{LAGNEW}),  the axial weak current, introduced through the replacement $f^2(k) \tilde A_\mu \to f^2(k)  \tilde A_\mu  +{\cal A}_\mu$, coupling to the quantity $\partial^\mu\Phi$  allows the  identification of  the pion decay constant parameter $F_\pi$ with the parameter $F$:
\begin{equation}
F_\pi=\frac{I^l(M_0)\,M^{2}_{0}}{F}\,-\,\frac{2\,I^{l}_{6}(M_0)\,M^{2}_{0}}{F} \frac{2 G_2 M_0^2\,I^{l}_6(M_0)}{1\,+\,4\,G_2\,M_0^2\,I^{l}_{8}(M_0)}\equiv F.
\end{equation}
The pion mass parameter keeps its formal expression~:
\begin{equation}
M^{2}_{\pi}=\frac{m\,M_0}{G_1\,F^2_\pi}.
\end{equation}
As for the scalar field the rescaling parameter (similar to eq.(\ref{QUARKSCALAR})) becomes  $g^2_{0S}=(2 N_c N_f\,I^l_{2S}(M_0))^{-1}$ and the corresponding canonical  mass parameter is~:
\begin{equation}
M^2_\sigma=	g^2_{0S}\,\left(\frac{\partial^2 W}{\partial {\cal S}^2}\right)_{{\cal S}=M_0}=\frac{I^l_{28}(M_0)}{I^l_{2S}(M_0)}\,
\left(4\,M^{2}_{0}\,+\,\left(\frac{\tilde I_{2}}{I_{28}}\right)_{vac}M^{2}_{\pi}\right).\label{SIGMASSCAN}
\end{equation}
Notice that the results differ from the sharp cutoff case since we now have three different integrals $I^l_2, I^l_{26}, I^l_{28}$ in place of one $I_2$.

\smallskip
We have a priori four parameters, $G_1, G_2, \Lambda$ and the bare quark mass $m$. We use 
$$ \Lambda=1\, GeV, \quad m=3.5\, MeV, \quad G_1=7.8\, GeV^{-2}.$$
and we obtain for the vacuum quark mass at zero momentum and the quark condensate~:
$$M_0= 371\, MeV \quad \Rightarrow\quad\left\langle \bar q q\right\rangle= -(286\, MeV)^3. $$
The $G_2$ parameter constrained to reproduce to reproduce the VDM phenomenology is~:
$$(G_2)^{VDM}= \frac{g^{2}_{V}}{M^{2}_{V}}=\left(\frac{2.65}{0.770}\right)^2\,GeV^{-2}.$$
With this value we obtain for the pion parameters~: 
$$G_2=(G_2)^{VDM} \Rightarrow F_\pi=91.3 MeV, \quad M_\pi=141.3\, MeV.$$
However in  nuclear matter calculations we allow for a small variation of $G_2$~:
$$G_2=0.78\,(G_2)^{VDM} \Rightarrow F_\pi=93.6 MeV, \quad M_\pi=137.8\, MeV.  $$
With this set of values the low momentum mass parameter for $M_\sigma$ defined in eq. (\ref{SIGMASSCAN}) and for other quantities are~: 
$$ M_\sigma=923\, MeV,\quad  M_V=1256\, MeV, \quad  M_A=1398\, MeV.$$ 
The numerical value of the vacuum scalar coupling constant is $g_{0S}=5.55$.
Notice that the numerical values of these mass parameters and the associated coupling constants are significantly altered by the fact that $I_{2S,V}$ differ from $I_2$. However the ratios $g_V/M_V$ and $g_{0S}/M_\sigma=0.006 \,MeV^{-1}$ which are the relevant quantity for nuclear matter calculation are not sensitive to this effect.

It is interesting to derive the potential of the equivalent linear sigma model.  It  is obtained through  a second order expansion in  
${\cal S}^2$ of the Dirac sea energy defined in eq. (\ref{HAT1}, \ref{HAT2}) around its vacuum expectation value $M^2_0$. We recover the usual linear sigma potential once we introduce a rescaled "`effective"' scalar field $(S)_{eff}=(F_\pi/M_0){\cal S}$,  normalized to $F_\pi$ in the vacuum~:
\begin{equation}
W^{L\sigma M}	= \frac{1}{4}\frac{(M^{2}_{\sigma})_{eff} - M^{2}_{\pi}}{2 F^{2}_{\pi}}
\left[(S^2)_{eff} -F^{2}_{\pi}\frac{(M^{2}_{\sigma})_{eff} - 3 M^{2}_{\pi}}{(M^{2}_{\sigma})_{eff} - M^{2}_{\pi}}\right]^2 -
F_\pi M^{2}_{\pi} (S)_{eff}.
\end{equation}
This potential has  the form of the linear sigma model potential and  the parameters,  instead of being arbitrary constants, have been dynamically generated. We remind that  our field $(S)_{eff}$ is a chiral invariant, so as to respect all chiral constraints for the mass evolution,  while the scalar field of the sigma model, $\sigma$ is not. The mass associated with this effective scalar field is~: 
$$(M^{2}_{\sigma})_{eff}=\left(4\,M^{2}_{0}\left(\frac{\tilde I_{2}}{I_{28}}\right)_{vac}\,+\,M^{2}_{\pi}\right)\quad
\Rightarrow\quad (M_{\sigma})_{eff}=659 \, MeV$$
 and the corresponding scalar coupling constant is the one of the quark level linear sigma model $(g_{0S})_{eff}=M_0/F_\pi$. Taking the nucleon as a naive juxtaposition of three constituent quarks, its mass evolution at low density goes as follows~:
 $$M^*_N\simeq M_N\, +\, 3 \,(g_{0S})_{eff}\,\big((S)_{eff}\, - \,F_\pi\big).$$	 

\smallskip
\paragraph{Completion of the model.}
In order to prepare the ground for a quark-diquark model of the nucleon (see next section), we also introduce an interaction in the quark-quark channel. We  limit ourselves to the  color $\bar{3}$,  scalar-isoscalar diquark channel. The corresponding contribution to the NJL interaction is
\begin{equation} 
{\cal L}_{diquark}=\frac{\tilde{G}_1}{2}\left(\bar{\psi}_c\,i\gamma_5\tau_2\beta_a\,\psi\right)\,
\left(\bar{\psi}\,i\gamma_5\tau_2\beta_a\,\psi_c\right)\label{LDIQUARK}
\end{equation}
where $\beta_a=\sqrt{3/2}\lambda_a$ ($a=2,5,7$) are color matrices, $\psi_c=i\gamma_2 \psi^{*}$ is the charge conjugate of the quark spinor and $\tilde{G}_1$ is a (positive) coupling constant. It can be generalized to  the delocalized version  exactly as for the case of the interaction in the $q\bar{q}$ channel. In the presence of the diquark channel, the bozonization procedure can be also done in presence of diquarks using the Nambu-Gorkov formalism. In the simplest approximation with a constant scalar background field, one obtains the mass and kinetic energy lagrangian for the scalar-isoscalar diquark fields, $\Delta^{a}_{S}$~:
\begin{equation} 
{\cal L}^l_{diquark}= \partial^{\mu}\Delta^{a}_{S}\,\partial_{\mu}\Delta^{a\dagger}_{S}\,-\,
M^{2}_{D\,}\Delta^{a}_{S}\,\Delta^{a\dagger}_{S}.
\end{equation} 
The diquark mass in a background scalar	field $\bar{\cal S}$ is: 
\begin{equation}
M^{2}_{D}(\bar{\cal S})=\frac{1}{2 N_c N_f \tilde{G}_1	I^l_2(\bar{\cal S})}\,-\,\frac{2 I^{l}_{1}(\bar{\cal S})}{I^l_2(\bar{\cal S})}.\label{DMASS}
\end{equation}

We now come come to the valence quark sector of the lagrangian. It also has an explicit delocalized form written below~:
\begin{eqnarray}
{\cal L}^{l}_{val}(x)&=& \bar{\Psi}(x)i\,\gamma^{\mu}\partial_\mu\,\Psi(x)\nonumber\\
&- & \int d^4x_1 d^4 x_2\,\bar{\Psi}(x_1)\,F_c (x_1-x)\left(\Sigma\,+
\,i\,P\,\gamma^5\,+\,\gamma^{\mu}(\tilde V_\mu\,+\,\gamma_5\,\tilde A_\mu)\right)(x)\,F_c (x-x_2)\,\Psi(x_2)\nonumber\\
&- &\int d^4x_1 d^4 x_2\,\bar{\Psi}_c(x_1)\,F_c (x_1-x)\,i\gamma_5\tau_2\beta_a\,\Delta^{a\dagger}_{S}(x)\,F_c (x-x_2)\,\Psi(x_2)\nonumber\\
&-&\int d^4x_1 d^4 x_2\,\bar{\Psi}(x_1)\,F_c (x_1-x)\,i\gamma_5\tau_2\beta_a\,\Delta^{a}_{S}(x)\,F_c (x-x_2)\,\Psi_c(x_2).
\end{eqnarray}
The full lagrangian, ${\cal L}^{l}_{mes}+{\cal L}^{l}_{diquark}+{\cal L}^{l}_{val}$,  can be utilized to describe the nucleon, generating models from a simple juxtaposition of constituent quarks to more refined ones  such as chiral solitons (nucleon bound by the chiral fields) or quark-diquark models including quark exchange diagrams. However with these models nuclear matter still remains unstable. This is the motivation for the next section where we introduce   on top of the effective NJL lagrangian  some confining interaction between quarks or between quark and diquark. If it is done the momentum of a valence quark inside the nucleon will be limited to $p\approx K^{1/2}_{string}\approx\Lambda_{QCD}\approx 200\,MeV$ which is much smaller than the scale $\Lambda\approx 1\, GeV$ entering the form factor. In such a case the delocalization effect is essentially not visible, and one can ignore the effect of the form factor  on the valence quark dynamics. If we again perform the chiral transformation on quark fields and vector fields, the valence quark sector lagrangian used in nucleon structure calculation can be safely taken as:
\begin{eqnarray}
{\cal L}_{val}&\simeq& \bar{Q}\left[i\,\gamma^{\mu}\partial_\mu\,-\,{\cal S}\,-\,\gamma^{\mu}\left( V_\mu\,+\,\gamma_5\, A_\mu\right)\right]Q\nonumber\\
& &-\,\bar{Q}_c\,i\gamma_5\tau_2\beta_a\,\Delta^{a\dagger}_{S}\,Q\,-\,
\bar{Q}_c\,i\gamma_5\tau_2\beta_a\,\Delta^{a}_{S}\,Q.
 \end{eqnarray}

\subsection{Concluding remarks on this section}
In the pure NJL picture, at finite baryonic density, the value of the constituent quark mass, which is the expectation value of the scalar field ${\cal S}$, is modified. It can be obtained self-consistently from a  gap equation modified by the presence of a Fermi sea. 
However in the real world baryonic matter is not made of independent constituent quarks but of clustered objects, the nucleons. These nucleons are embedded in the scalar background field, $\bar{\cal S}$, and the nuclear medium can be seen {\it a priori} as a shifted vacuum. The nucleon mass will depend in some way on the scalar background field and the energy density of  symmetric nuclear matter at the Hartree level reads~:
\begin{equation}
{E_0\over V}=\varepsilon_0=\int\,{4\,d^3 p\over (2\pi)^3} \,\Theta(p_F - p)\,\left(\sqrt{p^2+M^{2}_{N}(\bar{\cal S})}
\,-\,(M_N)_{vac}\right)
\,+\,W(\bar{\cal S})\,+\,9\,\frac{G_2}{2}\,\rho^2.\label{EOS}
\end{equation}
The expectation value for the scalar field is self-consistently obtained by minimization of the energy density,
\begin{equation}
	\frac{\partial\varepsilon_0}{\partial \bar{\cal S}}=0 \quad\Leftrightarrow\quad \frac{\bar{\cal S}- m}{G_1}=-2\,\left\langle \bar q q\right\rangle(\bar{\cal S})\,-\,\frac{\partial M_N}{\partial\bar{\cal S}}(\bar{\cal S})\,\rho^{N}_{s}(M_N(\bar{\cal S})),\label{MIN}
\end{equation}
which constitutes an in-medium modified gap equation. 
The connection between the field ${\cal S}$, normalized to the quark mass,  and the scalar field $s$ used in our previous work 
\cite{CEG01,CE05,CE07,MC08} is
$\bar{s}=(F_\pi/M_0)(\bar{\cal S}-M_0)$ and $\rho^{N}_{s}(M_N(\bar{\cal S}))$ is the nucleonic scalar density.
The scalar coupling constant of the nucleon to the effective scalar field (which is normalized to $F_\pi$ in the vacuum) is~:
$$(g_S)_{eff}(\bar{\cal S})=\frac{M_0}{F_\pi}\left(\frac{\partial M_N}{\partial \bar{\cal S}}\right)$$
which depends crucially on nucleon structure. For instance if the nucleon mass fully originates from confinement (bag  models), $\left(\frac{\partial M_N}{\partial \bar{\cal S}}\right) =0$, the scalar field just decouples from the nucleon, ($(g_S)_{eff}=0$). In this case  there is no shift of the vacuum and the scalar field is thus an irrelevant concept for nuclear matter studies. On the other extreme if the nucleon mass fully originates from chiral symmetry breaking (naive additive NJL, chiral soliton), then the nucleon mass in the medium is affected by the scalar field associated with the dropping of the chiral condensate. However in this case the attractive tadpole destroys stability. Only in the case where the nucleon mass has a mixed origin, the scalar background field can contribute to the nuclear attraction without destroying the stability and saturation properties. In that case 
 by rearranging its  quark structure linked to the confinement mechanism, the nucleon  reacts against the scalar field generating  effectively repulsive three-body forces. The origin of this repulsion lies in the decrease of the scalar coupling constant of the nucleon. In short a  possibly important   part of the saturation mechanism is associated with the progressive decoupling of the nucleon from the scalar field associated with the 
dropping of the chiral condensate. In the next section we will introduce nucleon models capable of achieving
the balance between large enough attraction and sufficient reaction. Of course, one falls here in the modeling
uncertainties. However we show below that a  stringent constraint exists for the numerical value of the
scalar nucleon coupling constant which is model dependent, from the value of the free nucleon sigma commutator.
 
The pion-nucleon sigma term is an important piece of experimental information. It is obtained from the Feynman-Hellman theorem~:
$\sigma_N=m \,(\partial M_N/\partial m) \simeq 50\,MeV$. It receives a contribution from the pion cloud, $\sigma_N^{(pion\,cloud)}$. According to previous works \cite{JTC92,BM92,LTY04,CE07} we expect:
$\sigma_N^{(pion\, cloud)}\simeq 20\, MeV$ which corresponds to a pion cloud self-energy of $- 420\,MeV$. For
the non pionic part an explicit calculation  in the NJL model shows  that the linear sigma model result is recovered  but with
the nucleon structure aspect hidden in the scalar coupling constant, $g_\sigma \equiv(g_S)_{eff}(M_0)$~:
\begin{equation}
\sigma^{(no pion)}_{N \sigma}=F_\pi g_\sigma\,\frac{M^{2}_{\pi}}{(M^{2}_{\sigma})_{eff}}.\label{SIGMANJL}
\end{equation}
Its  numerical value  has to be $\sigma^{(no pion)}_{N\sigma} = \sigma_N  -\sigma_N^{( pion\,cloud)} \simeq 50 - 20\simeq 30\, MeV$. This separation of the sigma term into two pieces is quantitatively supported by the lattice study of Leinweber et al \cite{LTY04}
on the nucleon mass evolution with the bare quark mass. In their work the pionic part of this evolution which has a
non-analytical behavior is calculated explicitly and subtracted out. For the rest  an expansion is made
in powers of $m_{\pi}^2$. The linear term in $m_{\pi}^2$ is linked to the non pionic sigma commutator, giving
a value  $\sigma_{N \sigma}^{(nopion)}\simeq 29\,MeV$, close to our value. With the value ${(M_{\sigma})_{eff}} = 659 MeV$ given previously it leads to $g_{\sigma }\simeq 7$.

Concerning the nucleon mass problem, there exist QCD sum rules which link in an approximate way the nucleon
mass to the condensate both for a free nucleon \cite{I83} and for a bound one   
\cite{CFG91,FKVW06}. For a dilute medium, these sum rules lead to the following mass evolution~: 
\begin{equation}
\frac{M^*_N}{M_N}\simeq \frac{\left\langle \bar{q}q\right\rangle}{\left\langle \bar{q}q \right\rangle_{vac}}=
1\,-\,\frac{\sigma_N\,\rho}{F^2_\pi\,M^{2}_{\pi}}=
1\,-\,\frac{\left(\sigma_{N\sigma}^{(nopion)}+\sigma_N^{(pion\,cloud)}\right) \,\rho  }{F^2_\pi\,M^{2}_\pi } \simeq
1\,+\,\frac{\bar{s}}{F_\pi}\,-\,\frac{\left\langle \Phi^2\right\rangle}{2\,F^{2}_\pi}\label{WRONG}
\end{equation}
where the last expression is the one obtained from the condensate evolution in the NJL model. We see that the 
quark condensate modification receives two contributions, one from the scalar field and one from the pion cloud, reconstituting at low density the
full pion nuclear sigma term. 
 
Our description brings important restrictions to this expression. Firstly the pion cloud contribution to the sigma commutator sigma commutator, 
$\sigma_N^{(pion\,cloud)}$,  contributes to the condensate evolution. It does not contribute to the mass evolution, otherwise chiral constraints would be violated (such as the presence of a term in $m_{\pi}$ in the NN potential, forbidden \cite{BK96} by chiral
symmetry). In fact its influence on the mass vanishes in the chiral limit and hence it is a small effect which we ignore. Only the scalar piece,  
$\sigma_{N \sigma}^{nopion}$, should then enter the mass evolution. In a pure chiral theory such as NJL the low density expansion of 
the mass evolution is then~: 
\begin{equation}
\frac{M^*_N}{M_N}\simeq
1\,-\frac{\sigma_{N\sigma}^{(nopion)}\rho}{F^2_\pi\,M^{2}_{\pi}}
\end{equation}
This separation already introduces a model dependence in the prediction of the mass evolution. This expression would hold for instance for an assembly of nucleons described as superpositions of constituent quarks or by chiral soliton models.
Secondly when we introduce confinement, {\it i.e}, when we go beyond NJL, the evolution of the nucleon mass with density, also depends  on the nucleon structure through the value of the coupling constant of the scalar field to the nucleon, $g_\sigma$. Indeed, to leading 
order, we have $M^*_N=M_N + g_\sigma\bar{s}$  with $\bar{s}=-g_\sigma\rho/(M^{2}_{\sigma})_{eff}$ which gives~: 
\begin{equation}
\frac{M^*_N}{M_N}\simeq  1\,-\,\frac{g_\sigma F_\pi}{M_N}\,\frac{\sigma_{N \sigma}^{(nopion)}\rho}{F^2_\pi\,M^{2}_{\pi}}\simeq
1\,-\,\frac{g_\sigma}{10}\,\frac{\sigma_{N\sigma}^{(nopion)}\rho}{F^2_\pi \,M^{2}_{}}
\end{equation}
In the linear sigma model where $g_\sigma=M_N/F_\pi=10$  we recover the Ioffe sum rule generalized at finite density just corrected from pionic effects. With confinement the value of the scalar coupling constant is reduced and the mass evolution is slower than the condensate one. The
suppression of the pionic contribution to the mass evolution further accentuates the difference between the mass and condensate evolutions. For instance at normal nuclear density the condensate has dropped by $\simeq 30\%$.  With the value $g_\sigma\simeq 7$ deduced
above, the mass reduction is significantly 
lower, $ \simeq 13\%$.

\section{Effect of confinement: simple  models for the in-medium nucleon} 

We now come to the last point of this paper, namely  the modeling of the nucleon mass origin and the scalar
response of the nucleon defined from the second derivative of the nucleon mass with respect to the scalar field~: 
\begin{equation}
\kappa_{NS}(\bar{\cal S})=\frac{\partial^2 M_N}{\partial \bar{s}^2}=\frac{M_0^2}{F_\pi^2}\left(\frac{\partial^2 M_N}{\partial \bar{\cal S}^2}\right)=\frac{M_0}{F_\pi}\left(\frac{\partial \left(g_S\right)_{eff}}{\partial \bar{\cal S}}\right).
\label{SCALARESP}
\end{equation}
For a nucleon made of the simple adjunction of three NJL constituent quarks (or a NJL quark and a NJL diquark) the scalar coupling constant is
independent of the scalar field and there is no  scalar response. The importance of the response is
 related to the respective roles of chiral symmetry breaking and confinement in the generation of the
 nucleon mass. 
In the following we will consider nucleons built from NJL quarks of mass M bound by some
 confining force. The information that we need is contained in the relation between the nucleon mass
 $M_N$ and the NJL mass $M$, $M_N = f(M)$. The scalar
 coupling constant of the nucleon is related to that of the NJL quarks, which is $M/F_\pi$, through
 $g_S=\frac{\partial f}{\partial M}\frac{\partial M}{\partial \bar s }$. The next derivative with respect
 to $\bar s$  gives the nucleon scalar response. A non vanishing value requires $\frac{\partial^2 f}{\partial^2 M}\neq 0$ 
and it  entirely arises from confinement.  For instance in soliton models where the quarks are bound only by chiral forces  the nucleon mass is proportional to the NJL mass and the second derivative vanishes. In the following we establish this relation for different models showing their effect on the saturation properties. 
 
In a previous work \cite{EC07} we have introduced a model of a nucleon made of three constituent
quarks bound together by a confining harmonic force. The magnitude of the scalar response which followed was 
too small to prevent the collapse of nuclear matter. We will come back later to this 
type of model. A possibility of improvement is to reduce the relative role of chiral symmetry breaking. This can
be achieved  by considering a nucleon made of a quark and a sufficiently light diquark to leave enough room for
confinement. A practical  advantage is that a three-body problem is transformed into a  simpler two-body problem. Beside this simplification,
there are theoretical and phenomenological reasons to favor a quark-diquark model of the nucleon with relatively
light scalar-isoscalar diquark. For instance the work of Shuryak {\it et al} on hadronic current-correlation 
functions based on a random instanton vacuum \cite{SH92} finds a strong attraction in the scalar-isoscalar 
channel leading to a diquark with a mass about $400\,MeV$. An axial-vector diquark is also found but with a
much larger mass of the order of $900\, MeV$. It is also possible to nicely reproduce the light baryon spectrum \cite{SA05} while a calculation without diquark correlations predicts an abundance of missing resonances \cite{CAP86}.

\begin{figure}[h]
\begin{center}
\includegraphics[scale=0.5,angle=0]{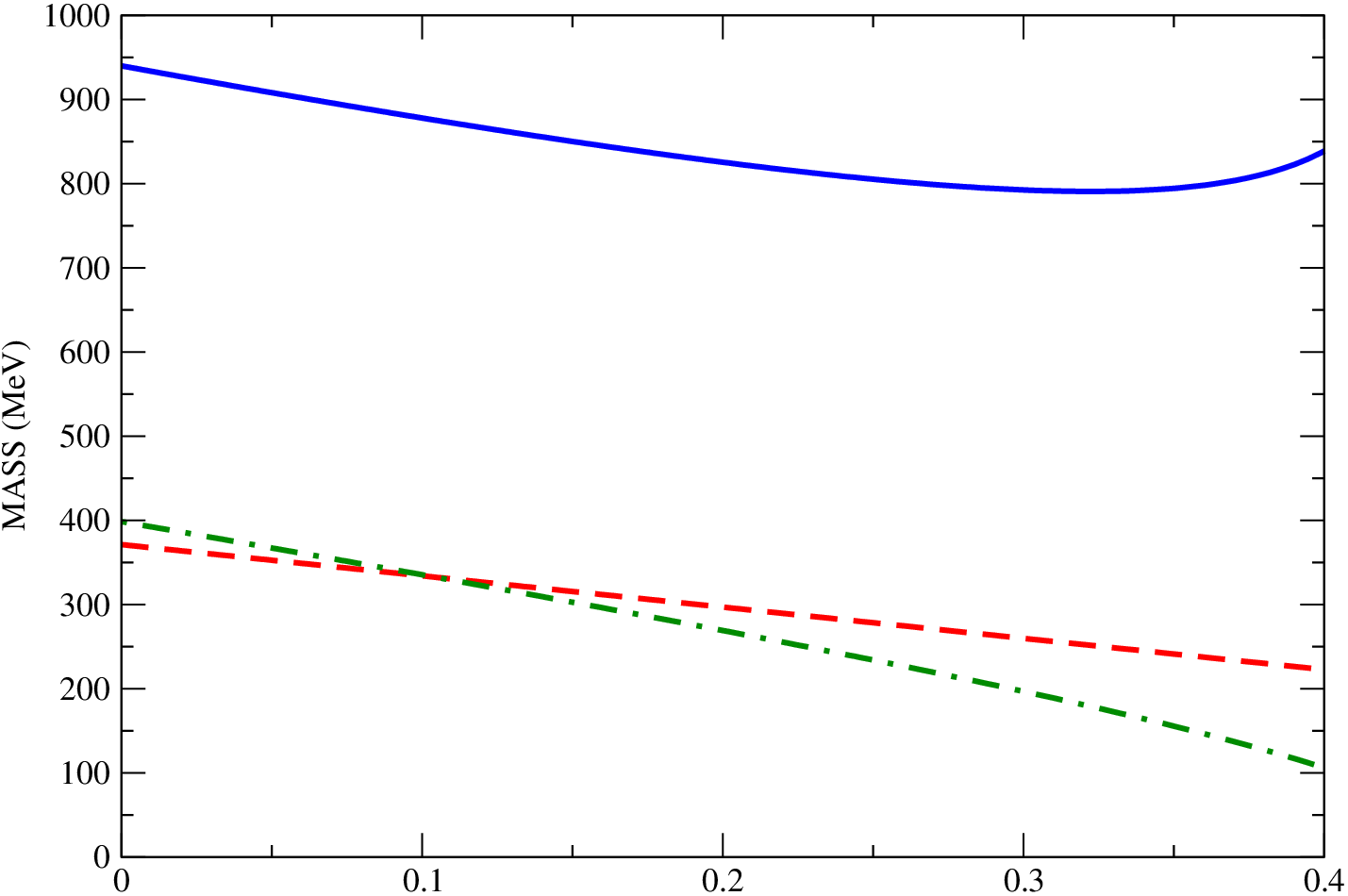}
\end{center}
\label{fig:mass}
\caption{Mass of the quark (dashed line), of the diquark (dot-dashed line) and of the nucleon (full line) versus the relative deviation, $\Phi=(M_0-M)/M_0\equiv \left|\bar{s}\right|/F_\pi$, of the scalar field with respect to its vacuum value.}
\end{figure}
As discussed in a set of works of Bentz {\it et al} (see ref. \cite{BT01} for application to nuclear matter),
it is possible from the NJL model to construct a nucleon with a diquark component. Introducing the  standard interaction in the diquark channel as discussed previously one obtains for the mass of the scalar diquark  result quoted in eq. (\ref{DMASS}). This mass is also medium dependent since it depends on the constituent quark mass. Its vacuum value  is strongly sensitive to the value of $\tilde{G}_1$ which defines the quark-quark interaction (see eq. \ref{LDIQUARK}). For $\tilde{G}_1=G_1$ it is exactly equal to the pion mass. Here  we choose
$$\tilde{G}_1=0.92\, G_1\quad \Rightarrow \quad M_D= 398.5\, MeV$$
which turns out to be nearly equal to the constituent quark mass in agreement with the work of ref. \cite{SH92}.  In \cite{BT01}, it was realized that to obtain  a scalar susceptibility,and consequently nuclear matter saturation  requires a confinement mechanism. An infrared cutoff $\mu_R\simeq\, 200 MeV$ was thus introduced in the Schwinger proper time regularization scheme. Such a prescription implies that  quarks cannot propagate at relative distance larger than $1/\mu_R$, hence mimicking a confinement mechanism. Here we propose to incorporate confinement in a more direct way. Since the diquark is in an anti-triplet color state, it is physically plausible that a string develops between the quark and the diquark as in a $Q\bar Q$ meson. We thus introduce a confining potential between the quark and the diquark~:
$$V(r) = \frac{1}{2}\,K\, r^2.$$
In the non relativistic limit, the problem reduces to solving the Schroedinger equation for a particle with reduced mass $\mu$, placed in an harmonic potential. In this limit the mass of the (in-medium) nucleon is given by~:
$$M_N(\bar{\cal S})=M(\bar{\cal S})\,+\, M_{D}(\bar{\cal S})\,+\,\frac{3}{2}\sqrt{\frac{K}{\mu(\bar{\cal S})}}\qquad\hbox{with}\qquad
\mu=\frac{M\, M_D}{M + M_D}.$$
We take for the string tension a standard value $K=(290 \, MeV)^3$. We obtain for the vacuum nucleon
mass $M_N=1304\, MeV$. The nucleon mass origin splits roughly into a chiral symmetry breaking component ($60\%$) and a confinement
component ($40\%$). The vacuum value scalar coupling constant of this nucleon to the effective scalar field is  $g_\sigma \equiv(g_S)_{eff}(M_0)=7.14$. This leads to the value of the non pionic piece of the sigma term~:  $\sigma^{(no pion)}_{N \sigma}= 30\, MeV$, as was required. In order to show that such a model is capable of describing the saturation properties of nuclear matter we calculate the energy of symmetric nuclear matter in the Hartree approximation, using eq.(\ref{EOS}, \ref{MIN}).
The resulting curve displays a saturation mechanism driven by the scalar nucleon response  ($\kappa_{NS}$, proportional to the second derivative of the nucleon mass with respect to $\bar{\cal S}$, eq. \ref{SCALARESP}) which has a positive value. Said differently the scalar coupling constant, $\partial M_N/\partial\bar{\cal S}$, is a decreasing function of $\left|\bar{s}\right|$ or the density. This translates into the fact that the nucleon mass stabilizes or even increases with increasing $\left|\bar{s}\right|$ (see fig. 1).  However the binding is nevertheless not sufficient unless we decrease artificially the vector coupling constant $G_2$ at a value much smaller than the VDM result. In order to improve the description, although this is not necessarily consistent with our present nucleon model, we add on top of the Hartree mean field result the pion loop (Fock term and correlation energy) contribution obtained in our previous work \cite{CE07}. Taking the value of $G_2$ at the value quoted previously, 
$G_2=0.78\,(G_2)^{VDM}$, we obtain a  decent saturation curve shown in fig. 2. Likely a fully consistent calculation 
within the model of the pion loop energy would  modify the result but a fine tuning on $G_2$ would be presumably 
sufficient to recover the correct saturation curve. The lesson of this simple model calculation seems to confirm our previous conclusions. The confinement effect (scalar response of the nucleon) is able  to stabilize nuclear matter and the pion loop correlation energy helps to get the correct binding energy.

\begin{figure}
\begin{center}
\includegraphics[scale=0.5,angle=0]{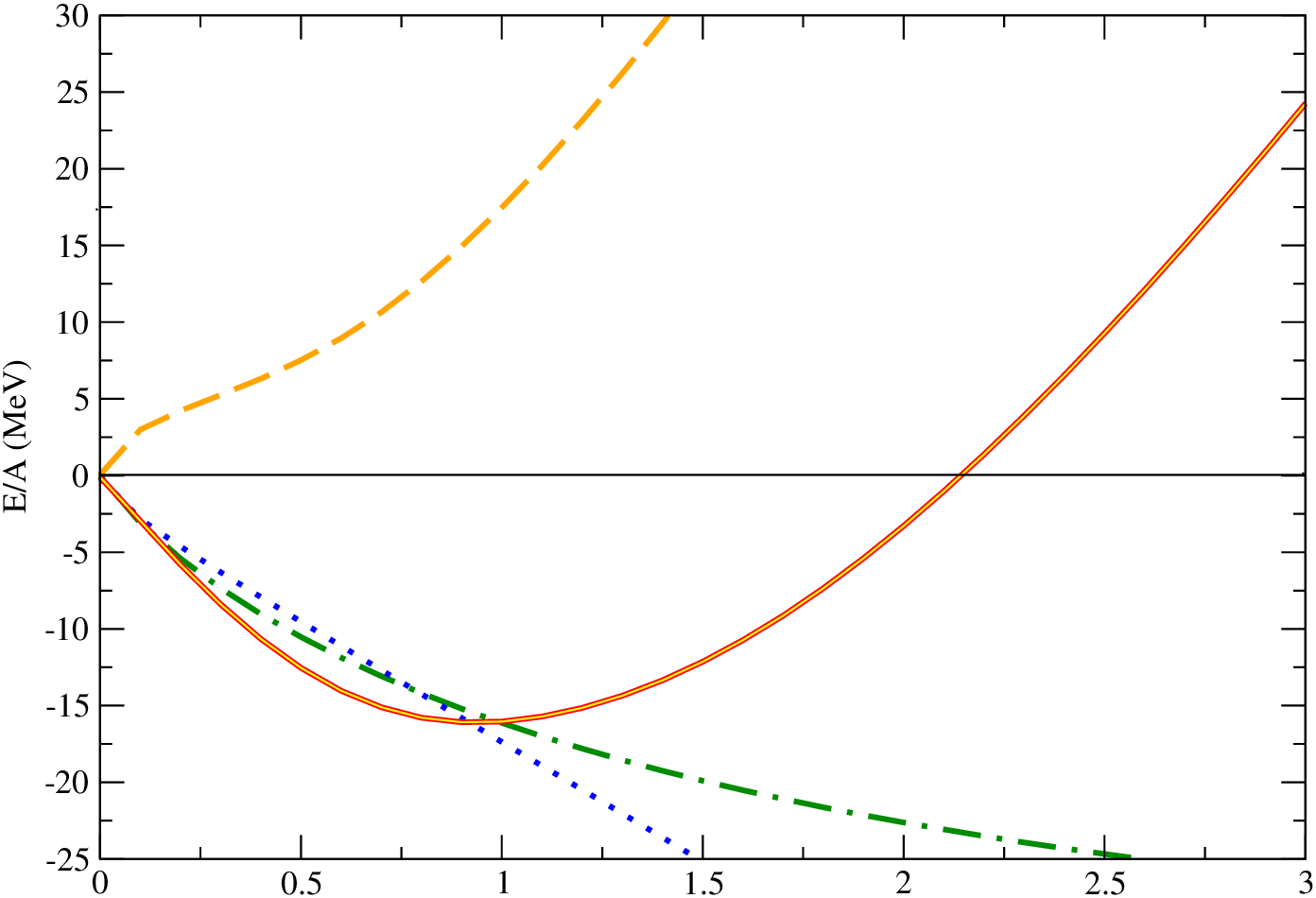}
\end{center}
\label{fig:eos}
\caption{Binding energy of nuclear matter versus nuclear matter density in units of normal density. 
The full line corresponds to the full result and the dashed line represents the  Hartree result. 
The dot-dashed line corresponds to  the contribution of the Fock term and the  dotted line  represents the
correlation energy. All the numerical inputs are given in the text.}
\end{figure}
\begin{figure}
\begin{center}
\includegraphics[scale=0.5,angle=0]{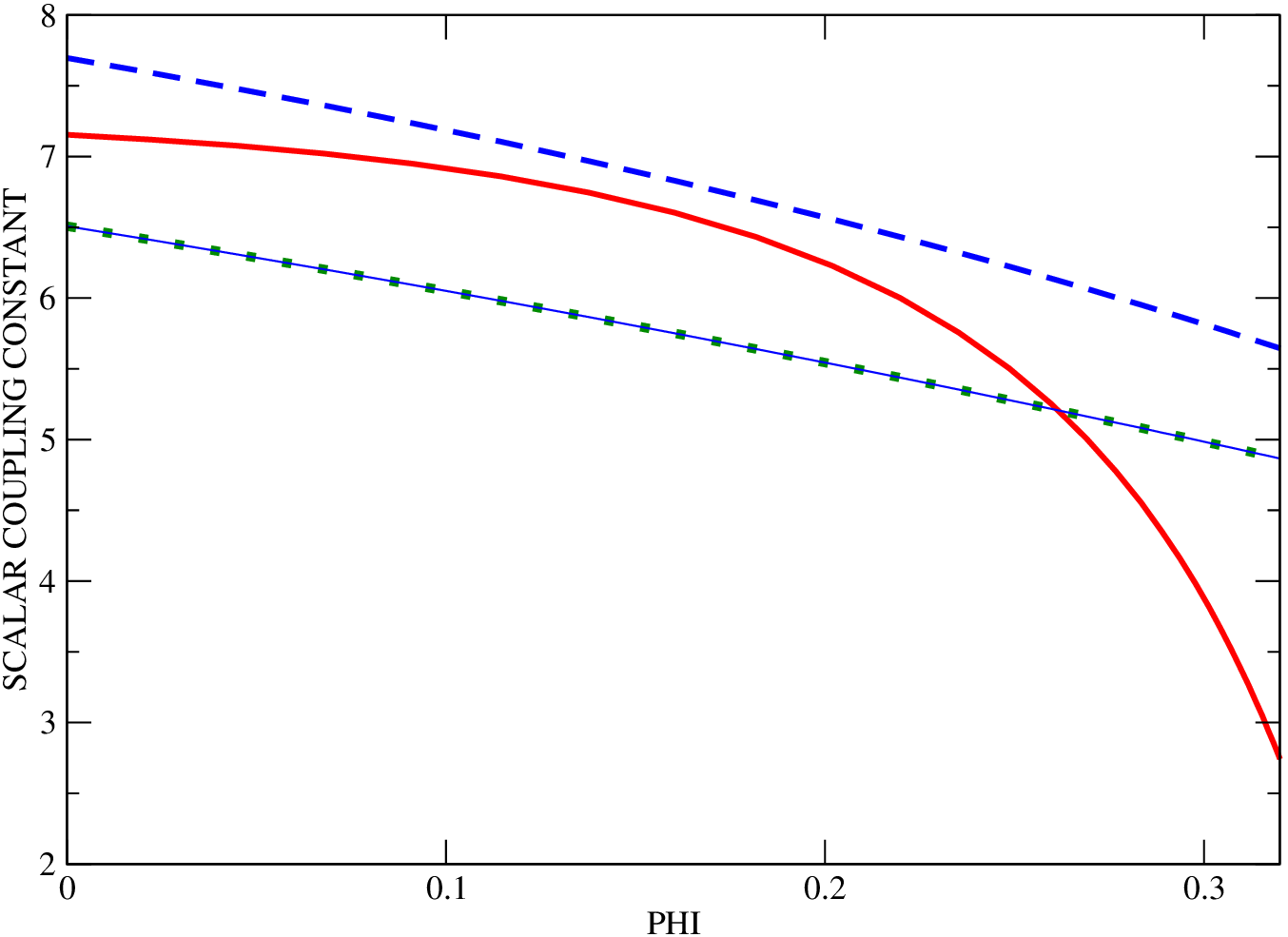}
\end{center}
\label{fig:scalarcoupl}
\caption{effective scalar coupling constant  versus the relative deviation, $(M_0-M)/M_0\equiv \left|\bar{s}\right|/F_\pi$, of the scalar field with respect to its vacuum value for the linear confining potential (dashed line), the quadratic linear potential (dotted line) and for the quark-diquark model (full line).}
\end{figure}

We have shown that an acceptable quark-diquark model of the nucleon makes plausible the role of the background scalar field in the nuclear binding. It is interesting to investigate if other confining mechanisms can achieve the same result. For this we have also studied models where the nucleon is made of three constituent quarks moving  in a mean-field linear confining potential but shifted with a constant attractive potential mimicking  short range attraction~:
$$V=\frac{1 + \gamma_0}{2}\left(K_2\, r\,-\,2\,V_0\right).$$
This model has been successfully utilized for baryon spectroscopy studies by Jena {\it et al} \cite{JBP97}. We do not aim to justify this particular equally mixed scalar and vector confining potentials, the main motivation being the existence of analytical solutions. 
 The energy of the lowest orbit, solution of the Dirac equation, is~: 
$$ E(M)=M\,-2\,V_0\,+\,\sqrt{K_2}\,x_q\,\quad\hbox{with}\,x_q\,\hbox{solution of}\quad x^{4}_{q}+ 2
\frac{M-V_0}{\sqrt{K_2}}x^{4}_{q}- (2.33811)^3=0$$
and the mass of the in-medium nucleon (in absence of CM correction) is $M_N(\bar{\cal S}=M)=3E(M)$.
Hence the quark mass contribution (essentially the chiral symmetry breaking contribution) to the quark orbital energy  and then to the nucleon is reduced due to the presence of the attractive shift, $-2 V_0$, leaving more room for the confining part. The scalar coupling constant (still omitting CM correction) can be written as
$$(g_S)_{eff}(\bar{\cal S})=\frac{M_0}{F_\pi}\left(\frac{\partial M_N}{\partial \bar{\cal S}}\right)\equiv
3 \,\frac{M_0}{F_\pi}\, q_s$$
where $q_s=\int d^3 r\,\left(u^2-v^2\right)(r)$ is the quark scalar charge. We see that the scalar field contribution to the sigma term is represented by the usual integrated scalar quark density as in bag models. In practice we also include in the numerical 
calculation the effect of CM correction using the results quoted in ref. \cite{JBP97}. If we  take $K_2 =( 300\, MeV)^2$ and  $V_0=200\,MeV$ it is possible to obtain a saturation curve but the saturation has the tendency to come too early. Certainly this point deserves a more detailed study. Here we wish to concentrate on the main result, namely a decreasing scalar coupling constant when increasing $\left|\bar{s}\right|$ as demonstrated by the dashed curve on fig. 3. We also checked that replacing the linear potential by a quadratic potential,
$$V=\frac{1 + \gamma_0}{2}\left(\frac{1}{2}\,K_3\, r^2\,-\,2\,V_0\right)\qquad\hbox{with}\qquad\ K_3 =( 300\, MeV)^3,\qquad V_0=200\,MeV,$$
one obtains similar results as depicted on fig. 3 (dotted curve). It is worthwhile to notice that this model differs from the one used in \cite{EC07} by the introduction of the constant attractive shift $-2\, V_0$. The energy of the lowest orbit is the solution of the equation~:
$$E=M\,-\,2\,V_0\,+\,\frac{3}{2}\sqrt{\frac{2\,K_3}{E\,+\,M}}.$$
Again this shift allows to reinforce the role of confinement in the origin of the nucleon mass. 

 Also shown on fig. 3 is the behavior of the scalar coupling constant 
for  the quark-diquark model. In this case, the decrease at low density is less strong which translates into a softer equation of state. According to a preliminary study based on a variational relativistic calculation the strong dropping beyond  $\left|\bar{s}\right|/F_\pi\approx 0.2$ (which roughly corresponds to normal density) might be to some extent an artifact of the non relativistic approximation.

\section{ Conclusion.}
We have studied the role played by the spontaneous breaking of chiral symmetry in the problem of the nuclear binding. The existence of a scalar field linked to the quark condensate emerges in chiral theories such as the NJL one. This field may be at the origin of the masses. This the case in the NJL model or the linear sigma one. In this case several things follow naturally. The partial restoration of chiral symmetry in dense matter implies a reduction in magnitude of the condensate and hence of the nucleonic mass, which could {\it a priori} account for the nuclear binding but in this case a tadpole term inherent in these theories destroys stability. A combination with the confining aspects is able to restore stability. Confinement indeed reduces the coupling constant of the scalar field to the nucleon and makes it field dependent. Equivalently it introduces a scalar response a the nucleon to this field in such a way that the nuclear medium reacts against a build up of the scalar field with increasing density, which helps in the saturation problem. However confinement should not be the only origin of the nucleon mass since in this case the scalar background field decouples from the nucleons. It is only in a mixed case, with a simultaneous influence of spontaneous symmetry breaking and confinement that the scalar field can be an efficient actor in the nuclear saturation problems. We have given examples of nucleonic models where this  balance is achieved. They require the role of confinement in the generation of the mass to be sufficient. We have shown how confinement affects the QCD sum rule for the
 in medium nucleon mass originally shown to follow the condensate evolution. Our  formula shows that the mass evolution is reduced as compared to the condensate one by a factor $r$,  ratio of the scalar coupling constants in the presence and in the absence of confinement. In addition, as we pointed out in previous works, only the non pionic part enters the mass evolution. These combined effects considerably reduce the mass evolution as compared to the condensate one. Nevertheless the remaining effect can be sufficient to make the scalar field of chiral symmetry breaking  an important actor in the nuclear binding question.


\begin{thebibliography}{99}
\bibitem{CE05} G. Chanfray and M. Ericson, EPJA 25 (2005) 151.
\bibitem{CE07} G. Chanfray and M. Ericson, Phys. Rev C75 (2007) 015206.
\bibitem{MC08}E. Massot, G. Chanfray, Phys. Rev C80 (2009) 015202.   
\bibitem{SW86} B.D. Serot, J.D. Walecka, Adv. Nucl. Phys. 16(1986) 1; Int. J. Mod. Phys. E16 (1997) 15.
\bibitem{CSWSX95} L.S. Celenza {\it et al}, Phys. Rev. C (2000) 035201, Annals of Physics 241 (1995) 1.
\bibitem{CWS01} L.S. Celenza, Huangsheng Wang and C.M. Shakin, Phys. Rev. C63 (2001) 025209.
\bibitem{G88} P.A.M. Guichon, Phys. Lett. B200 (1988) 235.
\bibitem{BT01} W. Bentz and A.W. Thomas, Nucl. Phys. A696 (2001) 138.
\bibitem{CEG01} G. Chanfray, M. Ericson, and P.A.M. Guichon, Phys.Rev. C63 (2001) 055202
\bibitem{CHAN86} L-H. Chan, Phys. Rev. Lett. 10 (1986) 1199.
\bibitem{RWB04} A. H. Rezaeian, N. R. Walet and  M. C. Birse, Phys.Rev. C70 (2004) 065203. 
\bibitem{LTY04} D.B. Leinweber, A.W. Thomas and R.D. Young, Phys. Rev. Lett 92 (2004) 242002.
\bibitem{JTC92} I. Jameson, A.W. Thomas and G. Chanfray, J. Phys. G18, L159 (1992).
\bibitem{BM92} M.C. Birse and J.E. McGovern, Phys. Lett. B292, 242 (1992). Progr. 
Theor. Phys. Suppl. 156, 124 (2004); nucl-th/0411014. 
\bibitem{I83} B. L. Ioffe,  Z. Phys. C18 (1983) 67
\bibitem{CFG91} T.D. Cohen, R.J. Furnstahl and D.K. Griegel, Phys. Rev. Lett. 67 (1991) 961.
\bibitem{FKVW06} P. Finelli, N. Kaiser, D. Vretenar and W. Weise, Nucl. Phys. A770 (2006) 1.
\bibitem{BK96} M.C. Birse and B. Krippa, Phys. Lett B381 (1996) 397.
\bibitem{EC07} M. Ericson and G. Chanfray, EPJA 34 (2007) 215.
\bibitem{SH92} T. Schafer, E.V. Shuryak and J.J.M. Verbaarschot, Nucl. Phys. B412 (1994) 143. 
\bibitem{SA05} E. Santopinto, Phys. Rev. C72 (2005) 022201(R).
\bibitem{CAP86} S. Capstick and N. Isgur, Phys. Rev. D34 (1986) 2809.
\bibitem{JBP97} S.N. Jena, M.R. Behera and S. Panda, Phys. Rev. D55 (1997) 291.




\end{thebibliography}
\end{document}